%% file: mandra_arxiv.tex
\documentclass[12pt]{article}

\usepackage{makeidx}
\usepackage{amsmath}
\usepackage{amsfonts}
\usepackage{amssymb}
\usepackage{graphicx}
\usepackage{authblk}

\usepackage{cite}
\usepackage{epstopdf}

\def\A{\mathcal{A}}
\def\S{\mathcal{S}}
\def\vs{\vec{\sigma}}
\def\vt{\vec{\tau}}
\def\N{\mathcal{N}}
\def\No{\N_0}
\def\p{\mathfrak{p}}
\def\T{\mathcal{T}}

\def\part#1{\left(#1\right)}
\def\parq#1{\left[#1\right]}
\def\parg#1{\left\{#1\right\}}

\def\pim{\pi_M(d\theta)}
\def\pimi{\pi_M(d\theta_i)}
\def\pinf{\pi_\infty(dt)}

\def\dirac#1{\delta\left(#1\right)_{\,(\!\!\!\!\mod 2)}}
\def\ave#1{\left\langle#1\right\rangle}

\begin{document}

\author[1,2,3]{Salvatore Mandr\`a}
\author[1,4,5]{Marco Cosentino Lagomarsino}
\author[1,2]{Bruno Bassetti}
\affil[1]{\small{Universit\`a degli Studi di Milano, Dip. Fisica, Milano, Italy}}
\affil[2]{INFN, Milano, Italy} 
\affil[3]{Universitat de Barcelona, Dep. F\'isica Fonamental, Barcelona, Spain}
\affil[4]{G\'enophysique / Genomic Physics Group, FRE 3214 CNRS "Microorganism Genomics"} 
\affil[5]{University Pierre et Marie Curie, 15, rue de l'\'Ecole de M\'edecine 75006 Paris France}

\date{}

\title{Typical rank of coin-toss power-law random matrices over
  $\mathbb{GF}(2)$}


\maketitle

\begin{abstract}
  Random linear systems over the Galois Field modulo $2$ have an
  interest in connection with problems ranging from computational
  optimization to complex networks. They are often approached using
  random matrices with Poisson-distributed or finite column/row-sums.
  This technical note considers the typical rank of random matrices
  belonging to a specific ensemble wich has genuinely power-law
  distributed column-sums. For this ensemble, we find a formula for
  calculating the typical rank in the limit of large matrices as a
  function of the power-law exponent and the shape of the matrix, and
  characterize its behavior through ``phase diagrams'' with varying
  model parameters.

\end{abstract}

\section{Introduction}

This technical note presents the calculation of the typical rank of
Boolean random matrices with power-law distributed column-sums. The
specificity of this calculation is that it applies to genuinely
power-law matrices, without finite cutoffs in the distribution. Before
presenting the results, we will give a brief description of the
context that motivates the calculation.\\

Random matrices with Boolean entries are often simple to treat, which
makes them important in many paradigmatic problems of different
branches of science.
For example, in computer science, they define the so-called random
\mbox{XOR-SAT} problem~\cite{dubois20023, monasson2007introduction,
  mezard2009information}, the simplest of an important class of
optimization problems at the interface of statistical
physics~\cite{mezard2003two, montanari2003nature} and computer
science~\cite{cheeseman1991really, selman1992new,
  mitchell1992hard}. The \mbox{XOR-SAT} problem consists in finding a
solution to the set of linear equations of $N$ Boolean variables and
$M$ equations $\A\vs = \vt$ over the Galois Field of order 2 (usually
indicated as $\mathbb{GF}(2)$), where the matrix $\A$ is extracted
from a prescribed ensemble of Boolean matrices.

The typical properties of the linear systems can be computed in the limit
of large matrices and fixed density of constraints $\gamma = M/N$.
For random matrices with constant row-sums (and thus
Poisson-distributed column-sums), the ``order parameter'' $\gamma$
plays a crucial role for the solution space of the corresponding
random \mbox{XOR-SAT} problem~\cite{mezard2003two}.
With increasing $\gamma$, the random \mbox{XOR-SAT} presents three
different regimes with some features of a 
thermodynamics phase~\cite{altarelli2008relationship}.
For $\gamma < \gamma_d$ a solution can be typically found by removing
iteratively all variables present in only one equation (trivial pivots
in the language of Gaussian elimination~\cite{mezard2003two,
  braunstein2002complexity}).  In this case, it can be shown that the
solution space is composed of only one cluster.  For $\gamma_d <
\gamma < \gamma_c$ matrices have typically a non-empty ``core'' (the
remaining part of the matrix after the recursive elimination of the
trivial pivots) and finding a solution requires a number of iterations
proportional to the cube of the size of the
core~\cite{braunstein2002complexity}.
Here, the solution space is split into many well separated
clusters. Finally, for $\gamma > \gamma_c$ in the typical case
solutions cannot be found (i.e. the solution space is empty).\\

In the field of complex networks, Boolean matrices are used to
represent empirical systems with many interacting agents: each agent
is labelled with an integer and the entry of the matrix $\A_{ij}$ is
equal to one only if agent $i$ interacts with agent $j$, and zero
otherwise. For instance, properties of the matrix $\A$ are useful to
control graph properties like hyperloops or critical sets of
independent nodes~\cite{kolchin1999random}.
In order to study the typical properties of such a system, it is
necessary to define an ensemble of matrices which conserves
characteristic properties of the empirical case. Of particular
interest are matrices with a power-law distribution of column-sums,
which are typical of many empirical graphs~\cite{albert1999internet,
  barabasi1999emergence,
  jeong2000large, guelzim2002topological}.\\

We have previously introduced a simple and analytically treatable
Boolean random matrix ensemble with a power-law distribution of the
column-sums $p(k) \sim k^{-\beta}$ and tunable
$\beta$~\cite{bassetti2007random,bassetti2007exchangeable}.
This paper describes an analytical approach to the problem of
the typical rank over $\mathbb{GF}(2)$ of random matrices belonging to
the this ensemble and compares the results to a numerical evaluation.
Previous approaches of this kind were applied to similar and more
sophisticated models, but were limited to distributions of the
row/column-sums with Poisson~\cite{mezard2003two} or
regular tails~\cite{franz2003replica},
or with power-law tails with a finite
cut-off~\cite{braunstein2002complexity, alamino2008typical}.

The calculation presented here is similar to the replica calculation
for spin-glasses~\cite{mezard1987spin}. It allows to find a formula
for the typical rank in the limit of large matrices as a function of
the model parameters $\gamma$ and $\beta$, which allows to derive
interesting phase diagrams.  In particular, we estimate a second order
transition in the typical rank varying the parameter $\gamma$.
We compares the results with the structure of solution space obtained
numerically. These results are resumed by interesting phase diagram
for the behavior of the linear system with varying density of
constraint $\gamma$ and power-law exponent $\beta$.

\section{Matrix Ensemble}\label{sec:matrix_ensemble}

This paragraph briefly describes the matrix ensemble. 
A more  exhaustive characterization can be
found in~\cite{bassetti2007random, bassetti2007exchangeable}.

\begin{figure}
	\centering
	\includegraphics[width=0.48\textwidth]{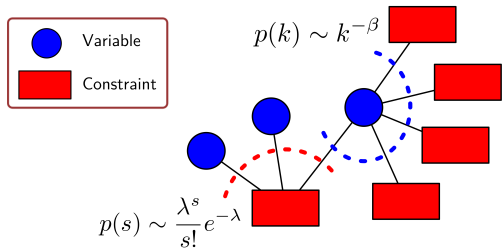}
	\caption{\label{fig:ex1} Schematic representation of the
          matrix ensemble. The probability that a variable is
          involved in $k$ constraints is asymptotically
          proportional to a power-law $p(k)\sim k^{-\beta}$ in the
          limit of large matrices.  Vice versa, the probability that a
          constraint contains $s$ variables is a
          Poisson distribution $p(s) \sim
          \frac{\lambda^s}{s!}e^{-\lambda}$, where $\lambda > 0$ is
          defined in the text.}
\end{figure}

The matrix ensemble (Fig.~\ref{fig:ex1}) was originally formulated as
a null model for (biological) transcriptional regulatory networks. It
is defined by the following generative algorithm.  For each column of
$\A$, (i) throw a bias from a prescribed probability distribution
$\pim$ and (ii) set the column elements of $\A$ to be $0$ or $1$
according to the toss of a coin with bias $\theta$. Since each column
is thrown independently, the resulting probability law is
\begin{equation}\label{eq:prob_definetti} 
	\p(\A) = \prod_{i=1}^N \int_0^1 \theta_i^{\sum_{j=1}^M
  	\A_{ij}}\left(1-\theta_i\right)^{\sum_{j=1}^M(1-\A_{ij})}\, \pimi.
\end{equation} 
Note that only columns are independent, while the row elements are
not independent, but symmetric by permutations.

To complete the model, one has to specify the choice for $\pim$, which
determines the behavior of the graph ensemble. To obtain a power-law
column-sums distribution we choose the two-parameter distribution
\begin{equation}\label{eq:pl} 
	\pi_M(d\theta) = Z^{-1}_M \theta^{-\beta}
	\chi_{\left(\frac{\alpha}{M},\ 1\right]} d\theta,
\end{equation}
where $\alpha > 0$ and $\beta > 1$ are free parameters,
$\chi_{\left(\frac{\alpha}{M},\ 1\right]}$ is the characteristic
function of the interval $\left(\frac{\alpha}{M},\ 1\right]$, taking
the value one inside the interval and zero everywhere else, and $Z_M =
\frac{(M/\alpha)^{\beta-1}-1}{\beta-1}$ is the normalization constant.
The function $\theta^{-\beta}$ of Eq.~\ref{eq:pl} gives a power-law
tail to the column-sums distribution.  Conversely, the cutoff on
$\theta$ defined by $\alpha$ poses a constraint on the number of nodes
with low degree, and will be used to control the probability to
extract a node with small $k$.  In the limit of large graphs (i.e. in
the limit $M,N\to\infty$, with $M/N = \gamma < \infty$) the
probability to extract a matrix with $k_i$ ones in the $i-th$ column
is asymptotically 
\begin{equation}\label{eq:prob_definetti2} 
	\p(\A) = \prod_{i=1}^\infty \int_0^\infty \frac{t_i^{k_i}\,e^{-t}}{k_i!}\,
	\pi_{\infty}(dt) \ , 
\end{equation} 
where 
\begin{equation}\label{eq:plinf} 
	\pi_{\infty}(dt) =
	(\beta-1)\alpha^{\beta-1}\chi_{[\alpha,\infty)} t^{-\beta}\ dt\,
\end{equation}
is the limit of the distribution in Eq.~\ref{eq:pl}. 
Eqs.~\ref{eq:prob_definetti2} and \ref{eq:plinf} imply that the probability
to have a column with $k$ ones and the probability to have a row with
$s$ ones in the limit of the large graphs are respectively $\p_c(k) =
\int_0^\infty \frac{t^k}{k!}e^{-t} \pi_\infty(dt) \approx k^{-\beta}$,
and $\p_r(s) = \frac{\lambda^s}{s!}e^{-\lambda}$, where $\lambda =
\gamma
\int_0^\infty t\,\pi_\infty(dt)$.\\

\begin{figure}
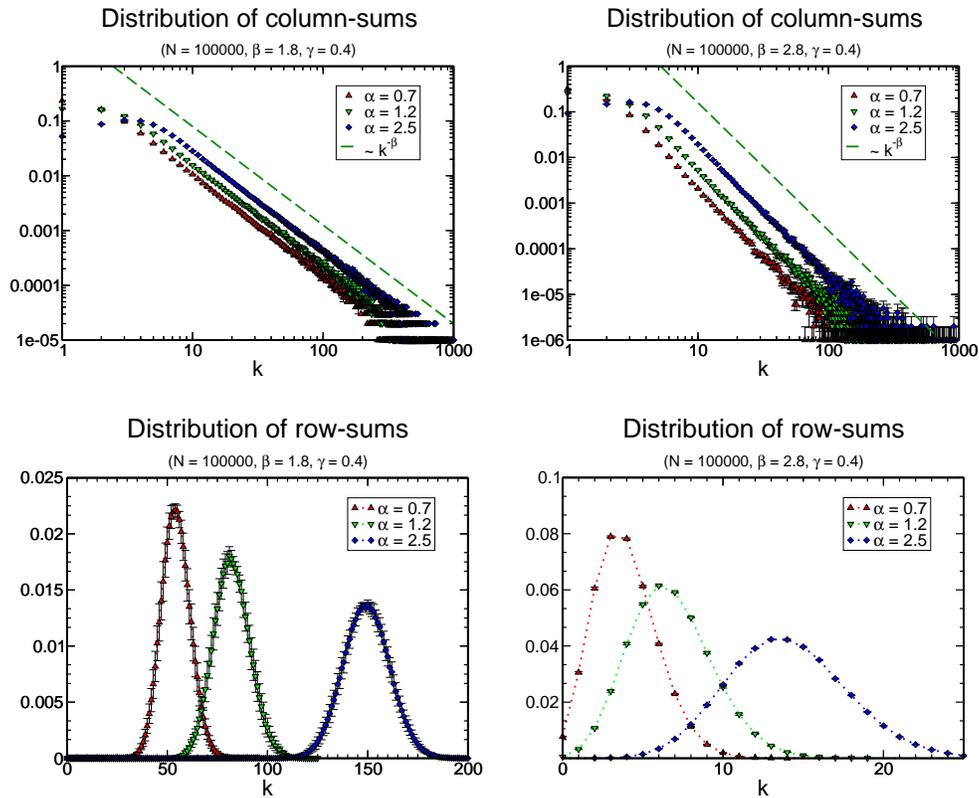

	\centering
	\includegraphics[width=0.45\textwidth]{graphs/distr_col2.eps}\ \ \ \ 
	\includegraphics[width=0.45\textwidth]{graphs/distr_col.eps}\\
	\vspace{0.5cm}
	\includegraphics[width=0.45\textwidth]{graphs/distr_row2.eps}\ \ \ \ 
	\includegraphics[width=0.43\textwidth]{graphs/distr_row.eps}
	\caption{\label{fig:distr_ones}  Distribution of
          nonzero entries for the matrix ensemble
          (Eq.~\ref{eq:prob_definetti2}), at $\beta = 1.8$ (left) and $\beta
          = 2.8$ (right).  As reported in the text, the column-sums
          (top) follow a distribution having a power-law tail with
          exponent $\beta$. The dashed (green) line is a guide to the
          eye. On the other hand, the distribution the row-sums
          (bottom) follows a Poisson-like distribution with mean
          depending on the value of the parameter $\beta$.}
\end{figure}

Fig.~\ref{fig:distr_ones} reports the distribution of the nonzero
entries of matrices extracted from the ensemble described by 
Eq.~\ref{eq:prob_definetti2}, for different values of $\alpha$ and
$\beta$.  As expected, the column-sums (top) follow a power-law
distribution while the distribution of row-sums (bottom) follow a
Poisson distribution. For $1 < \beta < 2$ the mean row-sum depends on
the dimension of the system as $\mu =
\frac{\beta-1}{\beta-1}\part{\frac{\alpha}{\gamma}}^{\beta-1}N^{2-\beta}$,
while for $\beta > 2$, the mean value of the distribution is 
independent of
the size of the system and it is $\mu =
\frac{\beta-2}{\beta-1}\frac{\alpha}{\gamma}$.

\section{Calculation of the Typical Rank}\label{sec:rank}

We will now consider the rank of a matrix belonging to the ensemble
described in the previous paragraph.  There are different methods for
computing the rank of a given matrix $\A$. Here, we exploit the
calculation of the number of solutions of the corresponding
homogeneous linear system
$$
	\N(\A) = \sum_{\vs} \dirac{\A\vs},
$$
where $\vs \in \parg{0,1}^N$ and
$\dirac{\vs}$ is different from zero only if $\vs \equiv
\vec 0\,(\textrm{mod}\,2)$. Since linear algebra applies, the number of
solutions of the homogeneous system over the finite field $\mathbb{GF}(2)$
can be expressed in terms of the dimension of the kernel of matrix
$\A$
$$
	\N(\A) = 2^{\textrm{null}(\A)}.
$$
Using the rank-nullity theorem
$$
	\text{rank}(\A) + \text{null}(\A) = N,
$$
the typical rank of random matrices will be
$$
	\ave{\text{rank}(\A)} = N - \ave{\log_2\N(\A)},
$$
where the average $\ave{\cdot}$ is carried over the matrix ensemble
in Eq.~\ref{eq:prob_definetti}.\\

In order to calculate the logarithm of the number of solutions we use
the known limit
\begin{align}\label{eq:repl}
\log \mathcal{X} = \lim_{k\to 0}\frac{\mathcal{X}^k - 1}{k},
\end{align}
where $\mathcal{X}$ is a generic random variable~\cite{mezard1987spin,
  derrida1981random}.  In principle, using the above limit, it is
possible to calculate the average $\ave{\log \mathcal{X}}$ knowing the
function $\phi(k) = \ave{\mathcal{X}^k}$, where $k$ is a real
parameter. However, the calculation of the function
$\phi(k)$ for any real $k$ is typically hard.

As proposed in~\cite{mezard1987spin, derrida1981random}, a feasible
protocol to compute $\phi(k)$ consists in calculating the $k-th$
moment of the random variable $\ave{\mathcal{X}^k}$ (i.e. evaluate
$\phi(k)$ for integer values of $k$) and then finding by interpolation
a reasonable extension for any real $k$.
In many cases~\cite{mezard1987spin, derrida1981random, franz2001exact,
  oppermann2005scaling} it is possible to find a well-behaved
extension of the function
$\phi(k)$, but this is not generally true~\cite{verbaarschot1985critique}.\\

Thus, we are interested in the calculation of the $k-th$ moment of the
number of solutions $\phi(k) = \ave{\N(\A)^k}$. As reported in Appendix 
\ref{sec:appendix_calc}, we find 
\begin{align}\label{eq:km4}
	\ave{\No^k(A)} 
	& = 2^{-kM} \sum_{\parg{m_\S}}\binom{M}{\vec{m}}\left[ \sum_{\T\in[k]} 
	\xi_M\left(\sum_{\S\in[k]} m_\S\, ]\S\cap\T[ \right)\right]^N,
\end{align}
where the sum is carried over $2^k$ integer variables labelled by an
element of $[k]$ (i.e. the set of all possible subsets of $\parg{1, 2,
  \ldots, k}$) constrained by $\sum_{\S\in[k]}m_\S = 1$ (``replica''
indices), and $]\Omega[$ is equal to one if the cardinality of the set
$\Omega$ is odd and zero otherwise. The function $\xi_M(h) := \int
\pim
\part{1-2\theta}^h$ is related to the moments of the column-sum
distribution of the random matrices extracted from the ensemble in
Eq.~\ref{eq:pl}.  Note that Eq.~$\ref{eq:km4}$ is not an approximation
but it is valid for any $M, N < \infty$.

In the limit $M\to\infty$ at fixed $x = h/M$ the function $\xi_M(h)$
can be written as
$$
	\xi(x) = \lim_{M\to\infty} \xi(h/M) = \int\pinf e^{-x\,t},
$$
where $\pinf = \lim_{M\to\infty} \pim$ (see Eq.~\ref{eq:plinf}). It is
immediate to observe that $\xi(x)$ is the moment-generating function
of Eq.~\ref{eq:prob_definetti}
$$
	 \p_c(k) = (-)^k k!\,\frac{d^k \xi(x)}{dx^k}.
$$
\\

Using the defining expression for the ensemble (Eq.~\ref{eq:prob_definetti}), 
Eq.~\ref{eq:km4} can be rewritten as
\begin{multline}\label{eq:km_limit}
  \ave{\No^k(\A)} \approx 2^{-km}\int[dx]\exp N \Bigg\lbrace\gamma 
  	\sum_{\S\in [k]}\mathfrak{S}(x_\S)\, + \\
	+\log\parq{\sum_{\T\in [k]}\xi_\infty
	\part{\sum_{\S\in [k]}]\S\cap\T[x_\S}} + o(1/N)\Bigg\rbrace,
\end{multline}
where the integration is carried over the rescaled variables $x_\S =
m_\S/N$ (with the constraint $\sum_{\S\in[k]}x_\S = 1$) and
$\sum_{\S\in[k]}\mathfrak{S}(x_{\mathcal{S}}) = \sum_{\S\in[k]}-x_\S\log x_\S$ is
the Shannon entropy. The above expression diverges exponentially with
the dimension $N$ of the matrices, and thus it is possible to use the
saddle point approximation. In order to compute the saddle point, it
is necessary to find the maximum of Eq.~\ref{eq:km_limit}
varying $x_\S$, i.e. it is necessary to solve a system of a $2^k$
variables for any integer $k$. Obviously, this is unfeasible and one
must impose a symmetry ansatz for the saddle point solution in order
to reduce the number of variables.

The simplest hypothesis it that the most symmetric solution would
dominate (in the theory of glassy systems this solution is usually
called replica symmetric (RS) solution)
\[
	\begin{array}{l}
		x_\emptyset = x\\
		x_\S = 1 - (2^k-1)x,\ \ \S\not=\emptyset,
	\end{array}
\]
where all variables are equal, except one in order to satisfy the
constraint $\sum_{\S\in[k]} x_\S = 1$. 
Here, the variable
$x$ plays the same role of the ``Edward-Anderson'' order
parameter in the Spin Glass theory \cite{mezard1987spin}: for
$x = 0$, the total entropy 
$\sum_{\S\in[k]}\mathfrak{S}(x_{\mathcal{S}})$ is exactly
zero and then only one state, i.e. the most symmetric state,
dominates the saddle point in Eq. \ref{eq:km_limit}. 
On the contrary, for $x = 1$, the total entropy assumes
the highest possible value and then many different states
contribute to the saddle point in Eq. \ref{eq:km_limit}.
\\

Using the RS ansatz, the asymptotic behaviour of the
$k-th$ moment of the number of solutions can be written as
\begin{multline}\label{eq:km_limit2}
  \frac{\log \overline{\No^k(\A)}}{N} = -k\gamma \log 2 
  	+ \max_{x} \Big\lbrace\gamma(2^k-1)\mathfrak{S}(x)\, +\\
  	+\, \gamma\mathfrak{S}(1-x) + 
	\log\parq{1 + (2^k - 1)\xi(2^{k-1}x)}\Big\rbrace.
\end{multline}
It is important to observe that the variable $k$ in 
Eq.~\ref{eq:km_limit2} can assume any real value and it can be considered
as a possible extension of the Eq.~\ref{eq:km4} in the limit of large
matrices. Eq.~\ref{eq:km_limit2} depends directly on the chosen
symmetry ansatz and is not guaranteed to be consistent. In our case,
we will show that Eq.~\ref{eq:km_limit2} gives results that agree
with numerical results.

\begin{figure}
	\centering
	\includegraphics[width=0.48\textwidth]{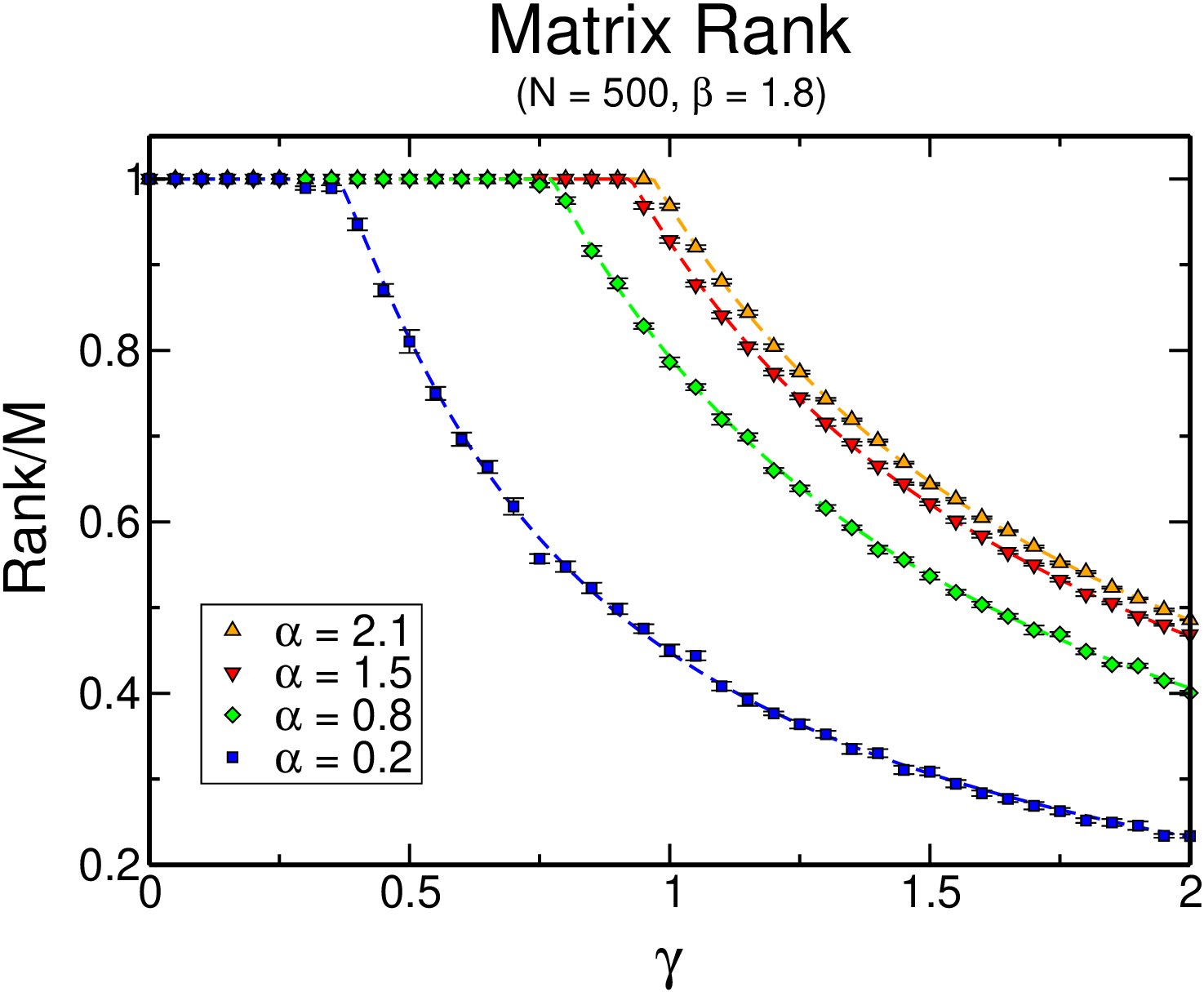}\ \ \ \ 
	\includegraphics[width=0.48\textwidth]{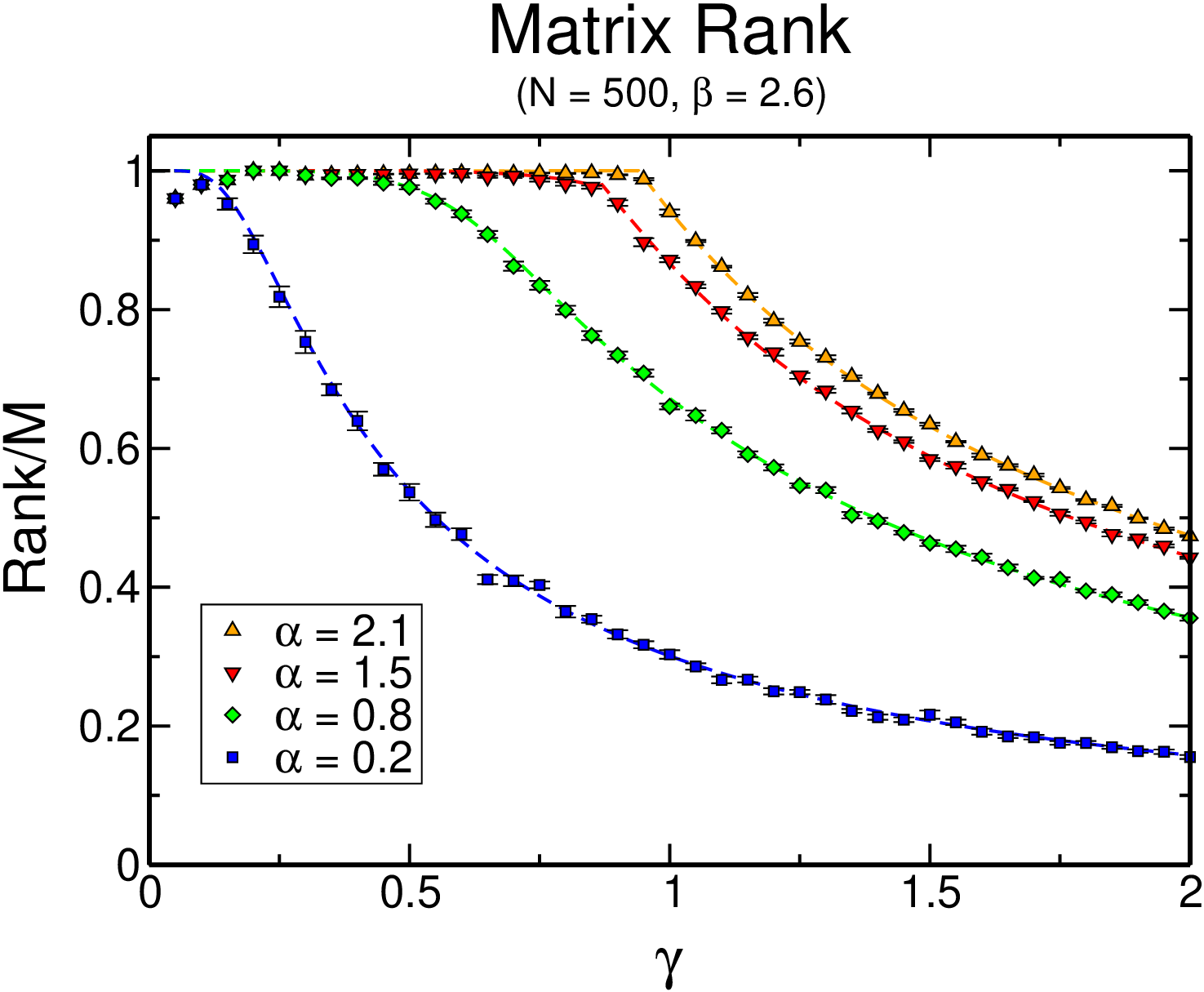}
	\caption{\label{fig:rank} Distribution of the typical rank
          obtained from simulation with $N = 500$ and $\beta = 1.8$
          (left) or $\beta = 2.6$ (right), varying the parameter
          $\gamma$. As shown in the figures, the numerical data are in
          agreement with the theoretical prediction obtained by
          Eq.~\ref{eq:km_limit3}. The deviation for small values of
          $\gamma$ is due to the small system size.}
	\vspace{1cm}
	\centering
	\includegraphics[width=0.48\textwidth]{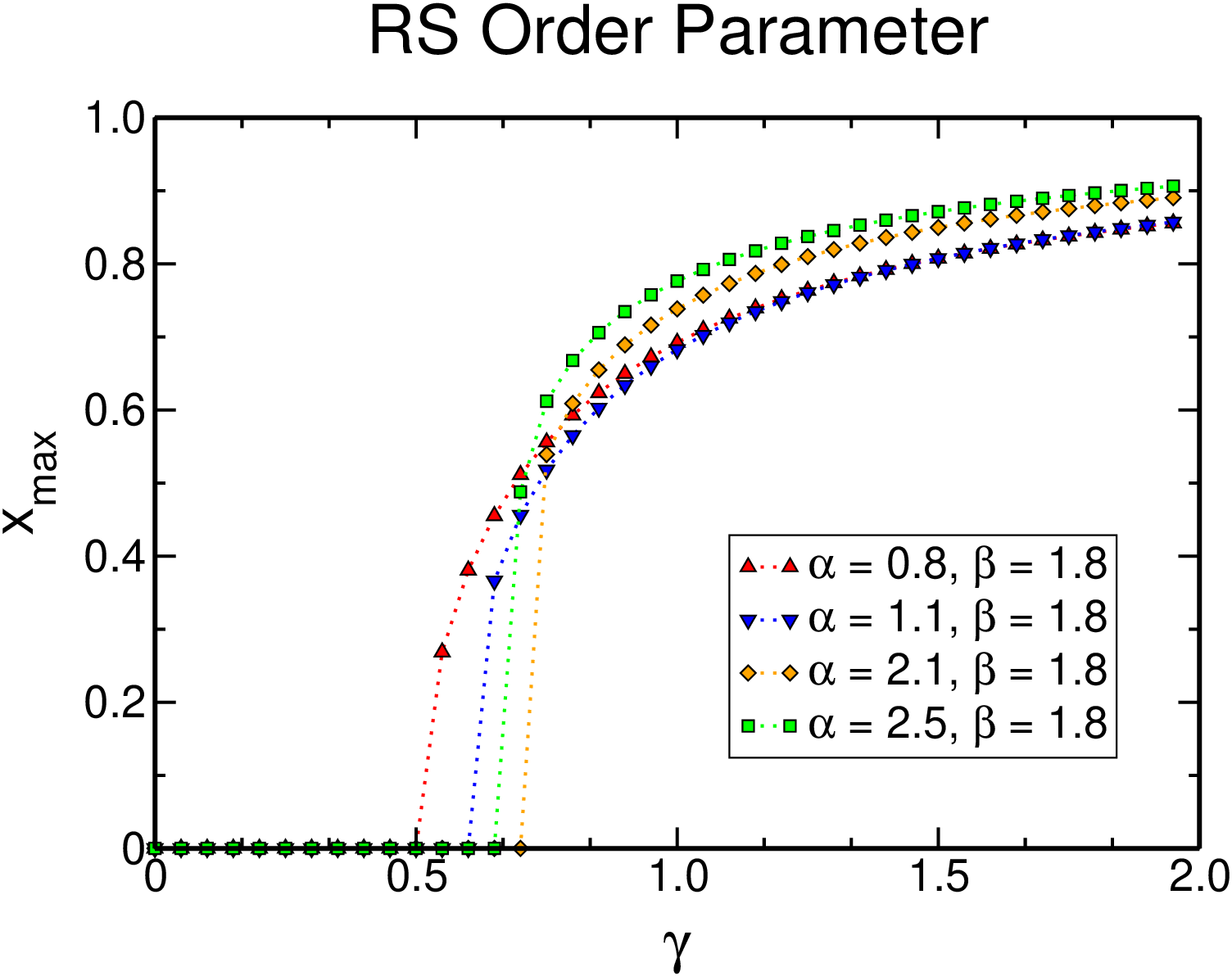}\ \ \ \ 
	\includegraphics[width=0.48\textwidth]{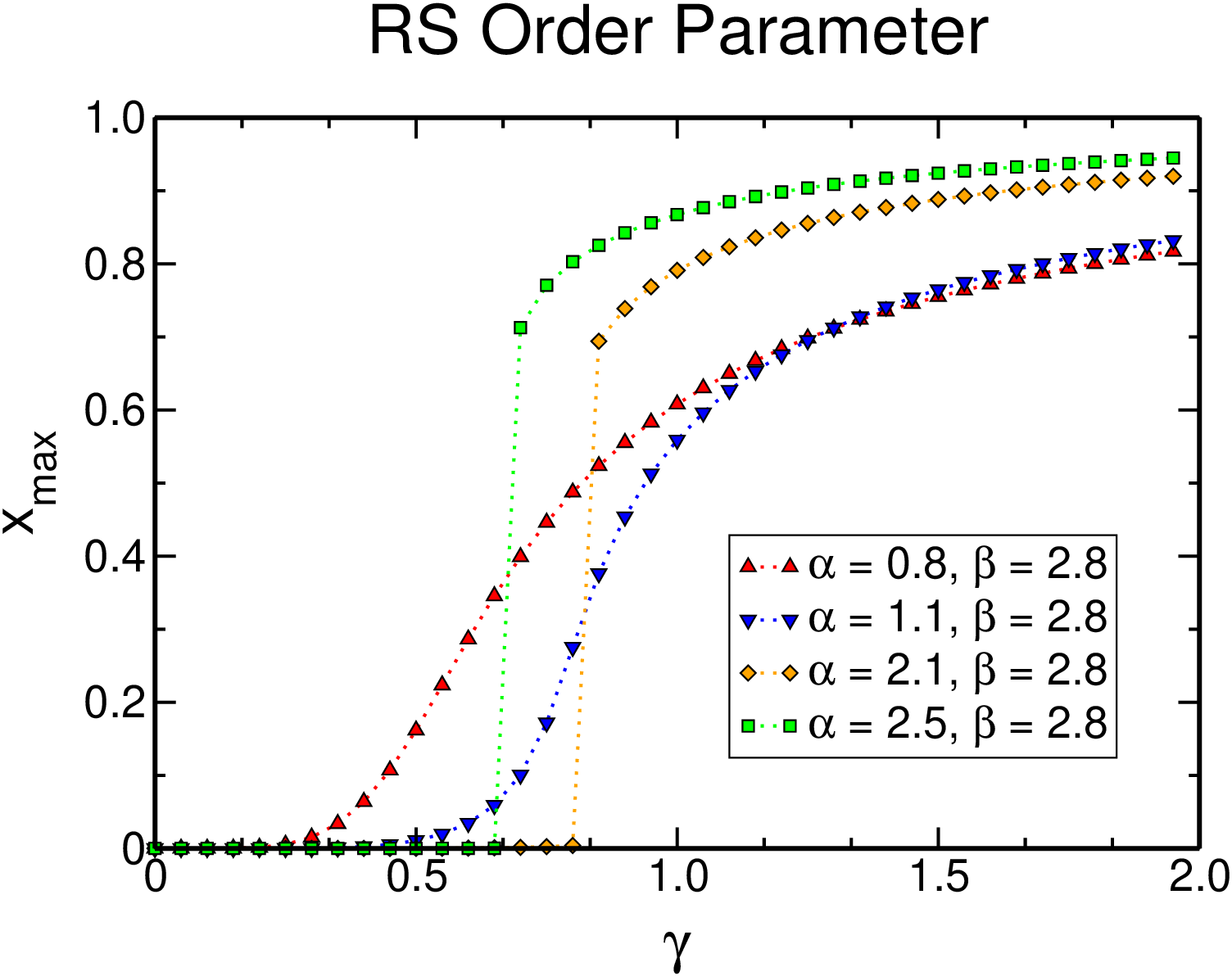}
	\caption{\label{fig:disc_x} Value of the 
          RS order parameter $x$ that
          maximizes Eq.~\ref{eq:km_limit3} at fixed $\beta$. For
          $\beta < 2$ (left), for any value of $\alpha$ there exists a
          critical value of $\gamma$ in which the value of $x$ at the
          maximum has a jump.  For $\beta \geq 2$ (right) and $\alpha$
          sufficiently small, the value $x_{max}$ does not have any
          discontinuity.  Otherwise, it is possible to identify a
          $\gamma_c$ (that depends on $\alpha$ and $\beta$) for which
          the value of $x_{max}$ has a jump.}
\end{figure}

We can now take the limit $k\to 0$. Thus we have
\begin{equation}\label{eq:km_limit3}
  \frac{\ave{rank(\A)}}{N} = 1 - \lim_{N\to\infty}\frac{\ave{\log_2 \No}}{N} = 
  \max_{x\in[0,1]}\parg{\gamma\mathfrak{S}_0(x) - \gamma + \xi\part{\frac{x}{2}}},
\end{equation}
where $\mathfrak{S}_0(x) = -x\log x + x$.  The above equation can be
used directly to find the typical rank of the matrices extracted from
the matrix ensemble proposed in Eq.~\ref{eq:prob_definetti}.
Fig.~\ref{fig:rank} compares the theoretical prediction of the typical
rank with simulations.  It is possible to observe that, independently
of the choice of the parameters $\alpha$ and $\beta$, the theoretical
prediction is in good agreement with the simulations.

Interestingly, the theoretical prediction of the rank
(Eq.~\ref{eq:km_limit3}) can have a second order discontinuity varying
the density of constraints $\gamma$, due to the fact that the value of
the RS order parameter 
$x$ which maximize the expression in Eq.~\ref{eq:km_limit3} can have a
jump (Fig.~\ref{fig:disc_x}).  In particular, we find that for any
$\beta > 2$ there exists a critical value $\alpha_c(\beta)$ such as
for $\alpha < \alpha_c(\beta)$ there are no jumps varying the
parameter $\gamma$. Instead, for $\alpha > \alpha_c(\beta)$, it is
possible to identify a critical value $\gamma_c(\beta)$ in which $x$
has a jump. On the contrary, for $1 <\beta < 2$ a discontinuity is
always present.

The presence of a second order discontinuity of the typical rank is a
signal of the fact that the totally symmetric solution (RS solution)
is no longer valid (even if it may still be a good approximation for
the calculation of the typical rank) caused by a spontaneous symmetry
breaking of the solution space in many well-separated
clusters~\cite{mezard1987spin}. In this case, a less symmetric
solution (called replica symmetry breaking (RSB) solution) dominates
the saddle point in Eq.~\ref{eq:km_limit}.  We did not explore
analytically this regime.

\section{Leaf Removal and Organization of the Solution Space}
\label{sec:leaf_removal}

As described in the previous paragraph, the typical rank of $\A$ is
related to the total number of solutions of the linear system $\A\vs =
\vec 0$. In particular, we found an analytical expression for the
typical rank which has sharp transitions when the parameters that
define the matrix ensemble vary continuously. As previously discussed,
these transitions are related to the clusterization of the solution
space.  This paragraph focuses on the geometrical organization of the
solution space of the linear system $\A\vs = \vt$ (the \mbox{XOR-SAT}
problem) and the comparison between numerical evaluations and our
theoretical predictions.  A general introduction to this problem can
be found
in~\cite{braunstein2002complexity, mezard2003two, mora2006geometrical}.\\

\begin{figure}
  \centering
	\includegraphics[width=0.80\textwidth]{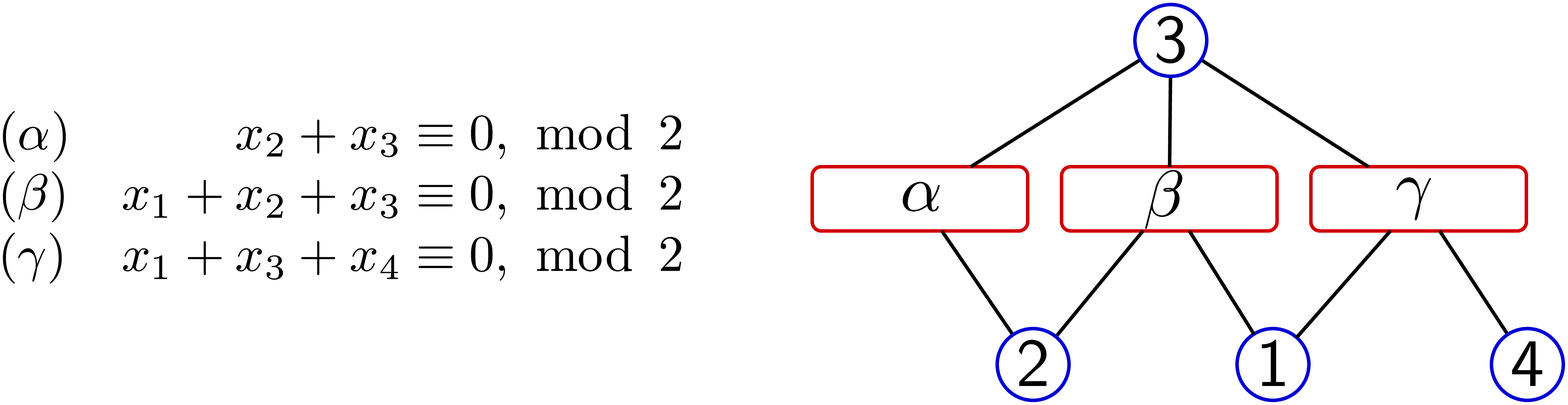}
	\caption{\label{fig:fg} Factor graph representation of the
	\mbox{XOR-SAT} problem. In the sketch the variables (columns) are
          represented by circles and the constraints (rows) by
          rectangles.}
\end{figure}

A system of linear equations in $\mathbb{GF}(2)$ can be conveniently
represented by factor graphs, defined by the matrix $\A$, in which
variables and constraints correspond to distinct types of nodes. If
the variable $i$ is present in the constraint $\alpha$, a link $(i,
\alpha)$ is drawn in the factor graph (Fig.~\ref{fig:fg}).

Following~\cite{braunstein2002complexity, mezard2003two,
  mora2006geometrical}, it is possible to obtain a precise definition
of clusters of solutions using the so-called ``leaf removal''
algorithm. The leaf removal algorithm is an iterative algorithm used
to gradually eliminate all trivially constrained variables (called
trivial pivots in the language of Gaussian elimination). It is easy to
prove that when a variable (called ``leaf'') is connected to only one
constraint, it is always possible to choose its value such that the
constraint is always satisfied (e.g. variable $4$ in
Fig.~\ref{fig:fg}).  The leaf removal algorithm is based on this
evidence and it is defined as follows: (i) pick a variable that
appears only in one constraint (leaf) and (ii) remove it together with
the only constraint it is connected to.  The process is iterated until
no leaves remain. The part of the factor graph that cannot be removed
by leaf removal iteration is called ``core'' and does not depend on
the order in which the leafs are removed.
In this case, the order parameter of the reduced linear system will be
\begin{equation}
	\gamma_{core} = \frac{M_{core}}{N_{core}},
\end{equation}
i.e. the density of constraints that are not trivially satisfied.

The presence of the core is related to the clusterization of the 
solution space. If $\gamma_{core} = 0$ (no core is present), the
problem to find a solution of the linear system $\A\vs = \vec \tau$ is
trivial (the complete solution can be found by running the leaf
removal in reverse direction, in a scheme usually called leaf
reconstruction) and the solution space is composed of only one
cluster. 
If $0 < \gamma_{core} < 1$, the core is not trivial (but not
over-constrained) and each solution of the linear system reduced to
the core variable defines a single cluster. 
All the solutions built
from a core solution by leaf reconstruction belong to the same
cluster. 
Finally, for $\gamma_{core} > 1$ the reduced linear system
for the core variables is over-constrained, so that no solutions are
typically found. 
\\

\begin{figure}
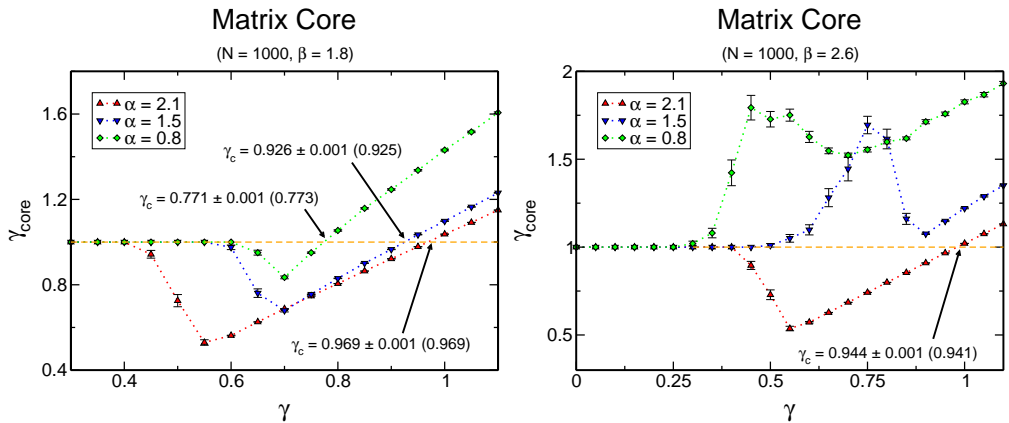

	\centering
	\includegraphics[width=0.48\textwidth]{graphs/core1.eps}
	\includegraphics[width=0.48\textwidth]{graphs/core2.eps}
	\caption{\label{fig:core}  Numerical simulation
          of the $\gamma_{core}$ varying the parameter $\gamma$, for
          different value of $\alpha$ and $\beta$. For $\beta < 2$
          (left), it is always possible to find the critical value
          $\gamma_c$ of inversion of the core. For $\beta > 2$
          (right), only for $\alpha$ sufficiently large it is possible
          to find the critical value $\gamma_c$.  In parenthesis
          the theoretical predictions.}
\end{figure}

\begin{figure}
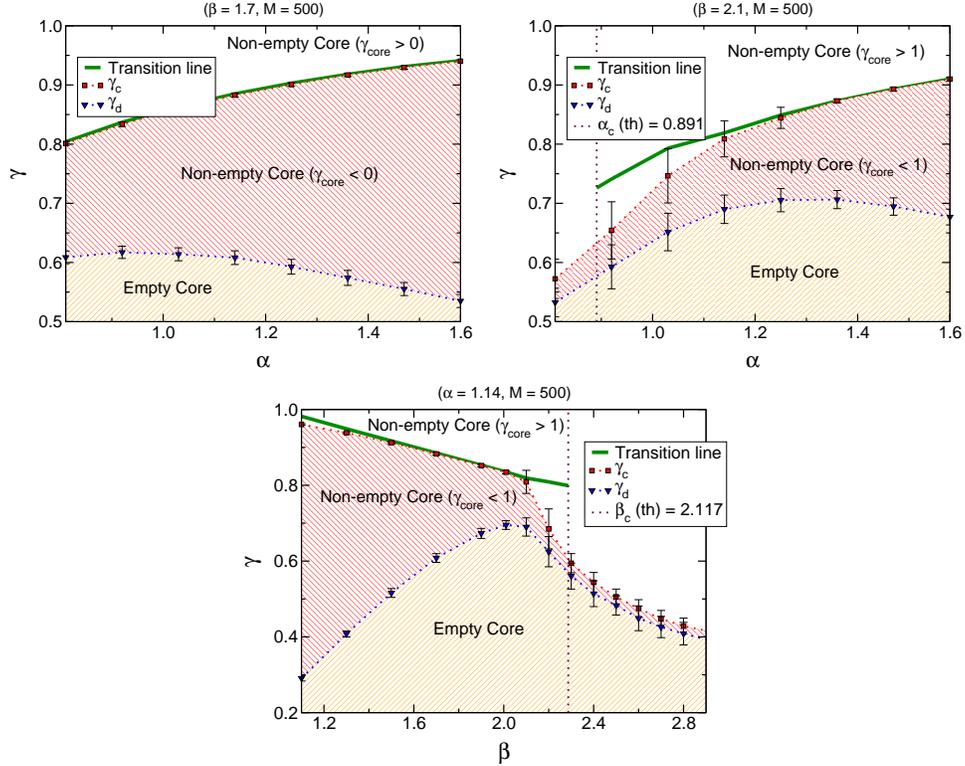

	\centering
	\includegraphics[width=0.45\textwidth]{graphs/trans_fase_b1-7_500.eps}\hspace{0.2cm}
	\includegraphics[width=0.45\textwidth]{graphs/trans_fase_b2-1_500.eps}\\\vspace{0.2cm}
	\includegraphics[width=0.47\textwidth]{graphs/trans_fase_a1-14_500.eps}
	\caption{\label{fig:trans} Phase diagrams obtained from
          simulation with $M = 500$ and $N = M/\gamma$. In the plots
          the results of the simulation are compared with the
          theoretical predictions (solid lines). The top panels
          contain fixed $\beta$ phase diagrams for $\beta>2$ and
          $\beta<2$, while the bottom panel is a fixed $\alpha$ phase
          diagram. The largest errors in the phase diagrams arise
          around the critical value $\alpha_c$ (top panels) and
          $\beta_c$ (bottom panel). This can be explained observing
          that near such critical values the matrix core is very small
          compared to the finite size fluctuations.}
\end{figure}

Fig.~\ref{fig:core} reports the curves of the typical
$\gamma_{core}$ varying the density of constraints $\gamma$
obtained by numerical simulations of the leaf removal
algorithm. As predicted in the previous paragraph, the presence of a non
over-constrained core depends on the choice of the parameter
$\beta$. For $\beta < 2$ (left panel), varying the parameter $\gamma$
it is always possible to identify three regimes: an empty core phase
($\gamma_{core}$ = 0), a non over-constrained core phase 
($\gamma_{core} < 1$) and an over-constrained core phase ($\gamma_{core} > 1$).
On the other hand, for $\beta > 2$ (right panel) the not
over-constrained ($\gamma_{core} < 1$) core is present only for
$\alpha$ sufficiently large.  All these results are resumed in the
phase diagrams obtained from in Fig.~\ref{fig:trans}.

\section{Conclusion}

In conclusion, we have presented a simple calculation of the typical
rank of random matrices with power-law distributed column-sums on the
Galois Field of order 2. The matrices can describe a graph or a sparse
linear system for Boolean variables. The calculation is based on a
fairly standard replica-like approach, where we compute the generic
$k$-th moment of the number of solutions of the associated linear
system and we consider the limit $k \to 0$ of its analytical extension
in the maximally symmetric case.

Differently from other models present in the
literature~\cite{mezard2003two,franz2003replica,braunstein2002complexity,alamino2008typical},
the simplicity of the matrix ensemble~\cite{bassetti2007random} that
we employ here allows to find an analytical expression for the typical
rank without having to impose any cutoff on the power-law
distribution.  As shown in Figs.~\ref{fig:rank}, the typical rank
calculated with our method is in fairly good agreement with the
numerical results.
We find that, as usually happens in this kind of
models~\cite{mezard2003two, braunstein2002complexity} the typical rank
can have a second order discontinuity with increasing density of
constraints $\gamma$.  This discontinuity is related to the
clusterization of the solution space in many well separated clusters
of the related XOR-SAT problem~\cite{mora2006geometrical}.  Our result
indicates that the same phenomenology can exist in presence of truly
power-law tails.

More in detail, since the matrix ensemble is defined as a function of
the model parameters $\alpha$, which sets a lower cutoff on the
row-sums and $\beta$, the exponent of the column-sum distribution, one
can study the variation of this threshold with ``phase diagrams''
where these parameters vary together with the density of constraints.
Specifically, the presence of the typical rank discontinuity at
$\gamma = \gamma_c$ depends on the choice of $\alpha$ and $\beta$. For
$\beta<2$ the discontinuity exists for any choice of $\alpha$.
Otherwise, for $\beta >2$, it is possible to identify a critical value
$\alpha_c(\beta)$ such that only for $\alpha>\alpha_c(\beta)$ a
critical value $\gamma_c$ exists. 

The role played by $\alpha$ in this model is similar to that played by
the constraint connectivity $K$ in the $K$--XOR-SAT
problem~\cite{monasson2007introduction}. In this case, the row-sum of
the matrix is equal to $K$, and the clustering of solution is possible
only if $K>2$. In our case, it is simple to verify that only for
$\beta<2$ the fraction of rows with two nonzero entries always
vanishes for every $\alpha$ in the large $N$ limit. Thus, we speculate
that the density of rows with two or less nonzero entries may become
important and affect in some cases the phase diagram for $\beta>2$,
causing the observed lack of the clusterization regime. 

Finally, the approach presented here, suitably generalized, may be
useful to study self-organizing properties of systems with many
interacting agents, where similar threshold phenomena can emerge as a
function of the properties of the network that defines the agent
interactions. In this case, the parameters of the matrix ensemble
represent tunable quantitative topological properties of the
interaction network such as the connectivity and the density of
interactions.
 
\appendix

\section{Calculation of {\boldmath $\ave{\No^k(\A)}$}}
\label{sec:appendix_calc}

\input{appendix.tex}


\bibliographystyle{unsrt}
\bibliography{referrer}

\end{document}

%% file: appendix.tex
\newcommand{\be}{\begin{equation}}
\newcommand{\ee}{\end{equation}}

\def\N{\mathcal{N}}
\def\Q{\mathcal{Q}}
\def\GF{\mathbb{GF}}
\def\A{\mathcal{A}}
\def\P{\mathcal{P}}

\def\vs{{\vec{\sigma}}}
\def\vt{{\vec{\tau}}}
\def\x{{\vec{x}}}

\def\dirac#1{\delta\left(#1\right)}
\def\parg#1{\left\lbrace#1\right\rbrace}
\def\part#1{\left(#1\right)}
\def\parq#1{\left[#1\right]}
\def\norma#1{\left|#1\right|}
\def\diracn#1{\,\mathbb{I}\!\part{#1}}
\def\E#1{\,\mathbb{E}\left[#1\right]}
\def\PHI{\Phi^{(k)}_{M,N}(\parg{m_\S})}

\def\S{\mathcal{S}}
\def\G{\mathcal{G}}
\def\D{\mathcal{D}}
\def\T{\mathcal{T}}
\def\N{\mathcal{N}}
\def\No{\mathcal{N}_0}
\def\NoA{\mathcal{N}_0(A)}
\def\xim{\xi_M}
\def\pim{\pi_M(d\theta)}
\def\pima{\pi_{M/\alpha}(d\theta)}
\def\pimi{\pi_M(d\theta_i)}
\def\p{\mathfrak{p}}

\def\X{\mathbb{X}}
\def\Ze{\mathbb{O}}

\def\bdirac#1{\delta\left(1,#1\right)}

In this appendix we explicitly calculate the $k-th$ moment of the number of solutions of the homogeneous linear system 
$\N_0(\A) = \sum_\vs \dirac{\A\vs}$, with \mbox{$\vs \in \parg{0,1}^N$} and $\dirac{\vs} = 1$ only if $\vs = 0$.
Let $\p(\A)$ a generic probability distribution for the random matrix $\A$: thus
the $k-th$ momentum can be written as 
$$
	\ave{\No^k(\A)} = \sum_{\X\in\parg{0,1}^N\otimes\parg{0,1}^k}\sum_{\A\in\parg{0,1}^M\otimes\parg{0,1}^N}\p(\A)
		\prod_{j = 1}^M \prod_{\alpha = 1}^k \dirac{\sum_{i = 1}^N \A_{ji}\X_{i\alpha}}.
$$
For simplicity, in the rest of the appendix we use the convention
\begin{align*}
	i\in\mathbb{N},\,i &= 1,\,\ldots,\, N \textrm{ (position\ of\ the\ row)}\\
	j\in\mathbb{N},\,i &= 1,\,\ldots,\, M \textrm{ (position\ of\ the\ column)}\\
	\alpha\in\mathbb{N},\,i &= 1,\,\ldots,\, k \textrm{ (number of the ``replica'')}\\
\end{align*}

Use probability distribution for our model (Eq. \ref{eq:prob_definetti}), 
the expression of the $k-th$ moment will be
\begin{align*}
	\ave{\No^k(\A)} &= \sum_\X\sum_A\p(\A)\prod_{j,\alpha}\frac{1+(-1)^{\sum_i \A_{ji}\X_{i\alpha}}}{2} =\\
	&= 2^{-kM} \sum_\X\sum_\A \int \parq{\prod_i \pimi} \cdot \\
		&\hspace{0.3cm}\cdot\parq{\prod_{j,\alpha}\part{1+(-1)^{\sum_i \A_{ji}\X_{i\alpha}}}}
		\parq{\prod_j \theta_i^{\sum_i \A_{ji}}\part{1-\theta_i}^{M-\sum_i \A_{ji}}},
\end{align*}
where we used the explicit representation of the Kronecker delta for binary variables
\[
	\dirac{\sigma} = \frac{1+(-1)^\sigma}{2}.
\]
At this level, it is possible to exchange the sums over $\A$ and the integration
to obtain
\begin{multline*}
	\ave{\No^k(\A)}
	= 2^{-kM} \sum_\X \int \parq{\prod_i \pimi} \cdot \\
		\cdot\prod_j \left[\sum_{\vec{a}\in\parg{0,1}^N}
		\prod_{\alpha}\part{1+(-1)^{\sum_i a_i\X_{i\alpha}}}
		\prod_i \theta_i^{a_i}\part{1-\theta_i}^{1-a_i}\right].
\end{multline*}
The last term does not depend explicitly on $j$ and then we above expression can
be rewritten as
\begin{multline*}
	\ave{\No^k(\A)}
	= 2^{-kM} \sum_\X \int \parq{\prod_i \pimi} \cdot\\
		\cdot\left[\sum_{\vec{a}\in\parg{0,1}^N}
		\prod_{\alpha}\part{1+(-1)^{\sum_i a_i\X_{i\alpha}}}
		\prod_i \theta_i^{a_i}\part{1-\theta_i}^{1-a_i}\right]^M.
\end{multline*}
Now, using the identity
\[
	\prod_{\alpha=1}^k\part{1+f(\alpha)} = \sum_{\S\subseteq[k]}\prod_{\alpha\in\S}f(\alpha),
\]
where $[k]$ is the set of all the possible subsets of $\parg{1,\ \dots,\ k}$, the above expression
becomes
\begin{multline*}
	\ave{\No^k(\A)} 
	= 2^{-kM} \sum_\X \int \parq{\prod_i \pimi} \cdot\\
		\cdot\left\lbrace\sum_{\S\subseteq[k]}\prod_i\left[\sum_{\sigma\in\parg{0,1}}
		\part{(-1)^{\sum_{\alpha\in\S}\X_{i\alpha}}\theta_i}^{\sigma}\part{1-\theta_i}^{1-\sigma}\right]\right\rbrace^M.
\end{multline*}
It easy to observe that the last term can be directly calculated. Thus, 
after a sum over $\sigma$ we obtain
\[
	\sum_{\sigma\in\parg{0,1}}
		\part{(-1)^{\sum_{\alpha\in\S}\X_{i\alpha}}\theta_i}^{\sigma}\part{1-\theta_i}^{1-\sigma} =
		1-2\theta_i\,{\bdirac{\sum_{\alpha\in\S}\X_{i\alpha}}},
\]
where $\bdirac{\sigma}$ equals 1 if and only if $\sigma = 1$, and
\begin{equation*}
	\ave{\No^k(\A)}
	= 2^{-kM} \sum_\X \int \parq{\prod_i \pimi} \left\lbrace\sum_{\S\subseteq[k]}\prod_i\left[
		1-2\theta_i\,{\bdirac{\sum_{\alpha\in\S}\X_{i\alpha}}}\right]\right\rbrace^M.
\end{equation*}
In order to complete the calculation, 
it is necessary to expand the term inside the curly brackets. 
Let $\parg{m_\S}$ the set of $2^k$ variables such that
$\sum_{\S\in[k]}m_\S = M$. Thus we have
\begin{multline*}
	\ave{\No^k(\A)}
	= 2^{-kM} \sum_\X \sum_{\parg{m_\S}}\binom{M}{\vec{m}}\cdot\\
		\cdot\prod_i \left\lbrace\int \pimi \prod_{\S\subseteq[k]}
		\left[1-2\theta_i\,{\bdirac{\sum_{\alpha\in\S}\X_{i\alpha}}}\right]^{m_\S}\right\rbrace,
\end{multline*}
where $\binom{M}{\vec{m}}$ is the multinomial. Using the simple identity
\[
		\left[1-2\theta_i\,{\dirac{\sum_{\alpha\in\S}\X_{i\alpha}}}\right]^{m_\S} =
		\left(1-2\theta_i\right)^{\dirac{\sum_{\alpha\in\S}\X_{i\alpha}}m_\S},
\]
we obtain 
\begin{equation*}
	\ave{\No^k(A)}
	= 2^{-kM} \sum_{\parg{m_\S}}\binom{M}{\vec{m}}\prod_i \left\lbrace\sum_{\X_i\in\parg{0,1}^k} 
	\xi_M\left(\sum_{\S\subseteq[k]}\dirac{\tilde{\X_i}(\S)}m_\S\right) \right\rbrace,
\end{equation*}
where we used the notation
\begin{align*}
	\tilde{\X_i}(\S) &:= \sum_{\alpha\in\S}\X_{i\alpha} \\
	\xi_M(h) &:= \int \pim \part{1-2\theta}^h.
\end{align*}
It is immediate to observe that the expression inside the curly bracket is independent on $i$:
\begin{align}\label{eq:km3}
	\ave{\No^k(A)} 
	&= 2^{-kM} \sum_{\parg{m_\S}}\binom{M}{\vec{m}}\left[ \sum_{\vec{x}\in\parg{0,1}^k} 
	\xi_M\left(\sum_{\S\subseteq[k]}\dirac{\tilde{x}(\S)}m_\S\right)\right]^N,
\end{align}
where $\tilde{x}(\S) = \sum_{\alpha\in\S}\vec{x}_{\alpha}$. The above expression can be
simplified if we define $\T$ as the set of the positions of the vector $\vec{x}$ 
different from zero. Indeed, the function $\tilde{x}(\S)$ can be expressed as
\[
	\tilde{x}(\S) = \left]\S\cap\T\right[
\]
where $\left]\Omega\right[=1$ if the cardinality of $\Omega$ is odd and zero otherwise. 
Thus, replacing the sum over $\vec{x}$ with the sum over $\sum_{\T\subseteq[k]}$ in 
Eq. \ref{eq:km3} we finally obtain
\begin{equation*}
	\ave{\No^k(A)} 
	= 2^{-kM} \sum_{\parg{m_\S}}\binom{M}{\vec{m}}\left[ \sum_{\T\subseteq[k]} 
	\xi_M\left(\sum_{\S\subseteq[k]} m_\S\, ]\S\cap\T[ \right)\right]^N.
\end{equation*}